\documentclass[11pt]{article}
\usepackage{a4}

\usepackage{amsmath}
\usepackage{amsthm}

\usepackage{epic,eepic}
\def\SetFigFont#1#2#3#4#5{}

\newtheorem{theorem}{Theorem}
\newtheorem{corollary}[theorem]{Corollary}

\def\paren#1{\left( #1 \right)}
\def\acc#1{\left\{ #1 \right\}}

\newcommand{\mr}[1]{{\cal M}({#1})}
\newcommand{\p}[1]{p({#1})}

\begin{document}
\bibliographystyle{alpha}
\title{On maximal repetitions of arbitrary exponent}

\author{Roman Kolpakov\thanks{Moscow University, Russia, {\tt foroman@mail.ru}}
\and Gregory Kucherov\thanks{CNRS (LIFL, Lille and J.-V.Poncelet Lab., Moscow) and INRIA Lille - Nord Europe, France, {\tt Gregory.Kucherov@lifl.fr}} 
\and Pascal Ochem\thanks{CNRS (LRI, Orsay and J.-V.Poncelet Lab., Moscow) France, {\tt Pascal.Ochem@lri.fr}}
}


\maketitle

\begin{abstract}
The first two authors have shown~\cite{KolpakovKucherovFCT99,KolpakovKucherovJDA00} that the sum the exponent (and thus the number) of
maximal repetitions of exponent at least 2 (also called runs) is linear in the length of the word.
The exponent 2 in the definition of a run may seem arbitrary. 
In this paper, we consider maximal repetitions of exponent strictly greater than 1.

\end{abstract}

\section{Introduction}

Repetitions (periodicities) are fundamental concepts in word
combinatorics \cite{Lothaire83,ChofKarh97handbook,KolpakovKucherovLothaire04}. Recall
that each word $w$ is characterized by the {\em minimal period}
$\p{w}$ and by the exponent $e(w)$ which is the ratio $\frac{\p{w}}{|w|}$. A
great deal of work in word combinatorics has been devoted to the study of
words that do not contain subwords of a given exponent
\cite{ChofKarh97handbook}. Another research direction, of more
algorithmic nature, is the efficient identification of all subwords of
a given exponent in a word \cite{KolpakovKucherovLothaire04}, which
raises the combinatorial question of the possible number of such
subwords. 

In \cite{KolpakovKucherovFCT99,KolpakovKucherovJDA00}, the first two
authors considered the notion of {\em maximal repetitions} of a word,
which are subword occurrences that cannot be extended outwards without
changing their minimal period. They proved that the number of maximal
repetitions of exponent at least $2$ is linearly bounded in the length
of word. It has been conjectured that this number is actually smaller
than the word length. It has been also proved that not only the number
of maximal repetitions of exponent $2$ or more is linearly bounded,
but the sum of exponents of these repetitions is linearly bounded
too. 
The linear bound on the number of repetitions, in turn, allowed them to prove that all such
maximal repetitions can be found in linear time. More recently, other
researchers attempted to improve these results by finding a simpler
proof of the linear bound implying a smaller multiplicative
constant. The last current achievement in this direction is presented
in \cite{CrochemoreIlieTintaCPM08}. 

A big question that remained open in this development concerns the
lower bound of $2$ on the exponent of considered repetitions. While
this bound is intuitively natural (as it requires some subword to be
consecutively repeated at least twice), it has no formal
justification. Moreover, word combinatorics provides many separation
results when the ``right'' bound on the exponent is not an
``intuitive'' number. For example, the famous Dejean's result states
that the exponents that can be avoided on a ternary alphabet are
exponents greater than $\frac{7}{4}$ \cite{Dejean72}. As another example, there are
exponentially many binary words avoiding exponents greater than
$\frac{7}{3}$, while there are only polynomially many of them avoiding
smaller exponents \cite{KarSha04}. 

In this paper, we completely lift the lower bound on the exponent and focus on the
maximal repetitions of {\em any} exponent greater than $1$. Note that
repetitions with exponent between $1$ and $2$ are subwords of the form
$uvu$ that can be viewed as {\em non-consecutive
  repetitions}. Therefore, in this paper we consider both consecutive
(periodicities) and non-consecutive repetitions. To the best of our
knowledge, the number of repetitions of exponent smaller than $2$ has
not been studied. 

Instead of directly counting the repetitions or the sum of their
exponents, we consider the {\em sum of exponents decremented by
  $1$}. The main idea is that repetitions with exponents close to $1$
(i.e. subwords $uvu$ with $|v|\gg |u|$) contribute to the sum with an
amount close to $0$. We prove that this sum is upper-bounded by
$n\ln(n)$ (Theorem~\ref{t1}) which immediately implies that the number
of maximal repetitions of {\em any} exponent greater than
$1+\varepsilon$ is bounded by $\frac{1}{\varepsilon}n\ln(n)$. On the
other hand, the number of {\em all} maximal repetitions can be
quadratic (Theorem~\ref{t2}). We also obtain that the lower bound for
the sum is $\frac{|w|}{k}-1$, where $k$ is the alphabet size, and we
characterize the word achieving this lower bound
(Theorem~\ref{t3}). Finally, we study this sum for the words
containing only repetitions with a period bounded by a constant. 

While the ``whole picture'' of the count of the number of maximal
repetitions with exponent smaller than $2$ is still incomplete, we
believe that our results represent the first step in this direction. 

\section{Definitions}
Recall that for any word $w$, the (minimal) {\em period}, denoted $\p{w}$ is
the minimal natural $p$ such that $w[i]=w[i+p]$ whenever positions $i$
and $i+p$ both exist in $w$.
The {\em exponent} of $w$ is defined as $e(w)=\frac{|w|}{\p{w}}$ ($|w|$ is
the length of $w$). 
A {\em root} of $w$ is any subword of $w$ of length $\p{w}$. The
prefix (resp. suffix) root of $w$ is the prefix (resp. suffix) of $w$
of length $\p{w}$.

Given $w$, a {\em maximal repetition} in $w$ is a subword $w[i..j]$
such that $\p{w[i..j]}>\p{w[i-1..j]}$ (provided that $i\neq 1$) and
$\p{w[i..j]}>\p{w[i..j+1]}$ (provided that $j\neq |w|$). Informally,
``maximality'' means that the subword is extended outwards as much
as possible so long as its period is preserved. 

In this paper, we will be interested in {\em maximal repetitions} of
any exponent greater than $1$. The set of these subwords of $w$ will
be denoted $\mr{w}$. 

Note that any two occurrences of the same letter in $w$ define a
maximal repetition with a period that is a divisor of the distance
between these occurrences. In this case, we will speak about a maximal repetition
{\em defined by} a letter match. 


\section{Sum of decremented exponents}
\label{sum}

For a word $w$, we will be interested in the sum of exponents of all
maximal repetitions, decremented by 1:
\begin{equation}
\sum_{r\in \mr{w}}\paren{e(r)-1}.
\label{main-sum}
\end{equation}
This quantity can be viewed as the difference between the sum of
exponents of all maximal repetitions and the number of these
repetitions. 

\begin{theorem}
\label{t1}
For every word $w$ of length $n$, we have $\sum_{r\in \mr{w}}\paren{e(r)-1}\le n\ln(n)$.
\end{theorem}
\begin{proof}
For each maximal repetition $r$ with period $p$, we distribute the value $e(r)-1=\frac{|r|-p}{p}$
over $(|r|-p)$ pairs of matching letters $(w[i],w[i+p])$, $w[i]=w[i+p]$
within the repetition. Each such pair contributes to the sum
with weight $\frac{1}{p}$. 
Consider two positions $i$ and $j$, $1\le i<j\le n$, in $w$. 
If $w[i]=w[j]$, then this match participates in some repetition, but it
is counted only if the period of this repetition is $j-i$, in which
case it
contributes to the sum with the amount 
$\frac{1}{j-i}$. 
We thus have $\sum_{r\in \mr{w}}\paren{e(r)-1}\le\sum_{1\le i<j\leq
  n}\frac{1}{j-i}=\sum_{i=1}^{n-1}\frac{n-i}{i}=n\sum_{i=1}^{n-1}\frac{1}{i}-(n-1)\le n\ln(n)$ for $n>2$.
\end{proof}

If we count only maximal repetitions of period at most $p$, then the
following bound holds. 
\begin{corollary}
\label{c1}
For every word $w$ of length $n$, we have $\sum_{r\in \mr{w},\p{r}\leq
p}\paren{e(r)-1}\le n(\ln(p)+1)$.
\end{corollary}
\begin{proof}
If only repetitions of period at most $p$ are considered, then,
according to the proof of Theorem~\ref{t1}, the sum is bounded as follows. $\sum_{r\in \mr{w},\p{r}\leq
p}\paren{e(r)-1}\leq \sum_{1\le i<j\leq \min\{i+p,n\}}\frac{1}{j-i}\le n(\ln(p)+1)$.
\end{proof}

Complementarily, if we count only maximal repetitions of period at
least $p$, then we get
\begin{corollary}
\label{c1a}
For every word $w$ of length $n$, we have $\sum_{r\in \mr{w},\p{r}\geq
p}\paren{e(r)-1}\le n\ln(n/p)$.
\end{corollary}
\begin{proof}
Similar to 
Corollary~\ref{c1}. 
\end{proof}

Assume now that we focus only on maximal repetitions of exponent
$(1+\varepsilon)$ or more, and we want to count their number. 
Theorem~\ref{t1} immediately provides a nontrivial upper bound. 

\begin{corollary}
\label{c2}
For every word $w$ of length $n$ and every $\varepsilon>0$, the number of maximal repetitions
of exponent at least $(1+\varepsilon)$ in $w$ is at most
$\frac{1}{\varepsilon}{n\ln(n)}$. 
\end{corollary}
\begin{proof}
Consider the sum of Theorem~\ref{t1}. Each repetition contributes at
least $\varepsilon$ to it and therefore the number of those is at most $\frac{n\ln(n)}{\varepsilon}$. 
\end{proof}

Similarly,
Corollaries ~\ref{c1} and \ref{c1a}
imply respective upper bounds $\frac{1}{\varepsilon}{n\ln(p)}$ and
$\frac{1}{\varepsilon}{n\ln(n/p)}$ on the 
number of maximal repetitions of exponent at least $(1+\varepsilon)$
and of period respectively at most $p$ and at least $p$. 

The following Theorem shows that the upper bound of Theorem~\ref{t1}
is asymptotically tight within a factor of 8 and that the number of
{\em all} repetitions of arbitrary exponent can be quadratic (to be
contrasted with Corollary \ref{c2}). 

\begin{theorem}
\label{t2}
Let $w=(0011)^{n/4}$. Then 
\begin{itemize}
\item[(i)] $\sum_{r\in \mr{w}}\paren{e(r)-1}\ge\frac{1}{8}n\ln(n)$.
\item[(ii)] the number of all maximal repetitions of $w$ is $\Theta(n^2)$,
\end{itemize}
\end{theorem}

\begin{proof}
(i) The whole word $w$ is an obvious repetition of period $4$, its contribution to
the sum 
is $(n/4-1)$. Any other repetition can be specified by a
match between two $0$'s or two $1$'s that occur at a distance other
than a multiple of $4$. 

Consider a repetition $r$ in which letter $0$ at some position $m$,
$m\equiv 1\pmod{4}$, 
matches letter $0$ at a position $\ell>m$, $\ell\equiv
2\pmod{4}$. This match corresponds to end letters of the repetition,
as $w[m-1]=1$ (if $m\neq 1$) while $w[\ell-1]=0$, and $w[\ell+1]=1$
(if $\ell\neq n$) while $w[m+1]=0$. 

Furthermore, this repetition has period $\ell-m=|r|-1$ and this period
is minimal, as word $w[m..\ell-1]$ contains one more $0$ than $1$'s
and therefore the number of $0$'s and the number of $1$'s in $w[m..\ell-1]$ are
mutually prime, which shows that $w[m..\ell-1]$ is primitive (i.e. not
an integer power of some other word). 

Therefore, any two such positions $m$ and $\ell$ define a repetition that contributes
$1/(\ell-m)$ to the sum. 
In total, all such repetitions contribute $\sum_{i=1}^{n/4} (n/4-i+1)/(4i-3) \geq \frac{1}{32}n\ln(n)$.

There are three other symmetric cases: one corresponds to another way
of matching two $0$'s and the other two correspond to matching two $1$'s. 
The four cases together yield
$\sum_{r\in \mr{w}}\paren{e(r)-1}\ge (\frac{n}{4}-1)+
4\frac{1}{32}n\ln(n)\geq \frac{1}{8}n\ln(n)$. 

(ii) is obvious from the above, as the number of pairs of $0$'s and
pairs of $1$'s defining repetitions is quadratic. 
\end{proof}

We now focus on the lower bound for sum (\ref{main-sum}). 
In the rest of the paper, we assume that we have a $k$-letter alphabet
$A_k=\acc{a_1,a_2,\dots,a_k}$. 

\begin{theorem}
\label{t3}
For all $w\in(A_k)^*$, $\sum_{r\in \mr{w}}\paren{e(r)-1}\ge\frac{n}{k}-1$ and
the equality holds
if and only if $w=\paren{a_1 a_2 \ldots a_k}^{\frac{n}{k}}$ (modulo a
permutation of alphabet letters). 
\end{theorem}
\begin{proof}
Given a word $w\in (A_k)^*$, consider all occurrences of a letter
$a_i\in A_k$ in $w$, and let $d^i_1,d^i_2,\ldots,d^i_{\ell_i}$
be the distances between all consecutive occurrences of $a_i$ in
$w$. Consider the sum 
\begin{equation}
\sum_{a_i\in A_k} \sum_{j=1}^{\ell_i}\frac{1}{d^i_{j}}. 
\label{eq1}
\end{equation}
Observe that $\sum_{r\in \mr{w}}\paren{e(r)-1}\ge \sum_{a_i\in
  A_k}\sum_{j=1}^{\ell_i}\frac{1}{d^i_{j}}$ since two consecutive
occurrences of $a_i$ necessarily participate in a repetition with 
period equal to the distance between these occurrences, and then
contribute to sum (\ref{main-sum}) (see proof of Theorem~\ref{t1}). 

Therefore, if we
construct a word that minimizes sum (\ref{eq1}) and for which
$\sum_{r\in \mr{w}}\paren{e(r)-1}= \sum_{a_i\in
  A_k}\sum_{j=1}^{\ell_i}\frac{1}{d^i_{j}}$, this will prove that this
word also minimizes 
sum (\ref{main-sum}). 
%
Our goal is to prove that this minimum is reached if and only if for
any letter $a_i$, all $d^i_j=k$, i.e. on words of the form
$w=\paren{a_1 a_2 \ldots a_k}^{\frac{n}{k}}$ (modulo a 
permutation of alphabet letters). Clearly, for such words, sum
(\ref{main-sum}) and sum (\ref{eq1}) are both equal to
$\frac{n}{k}-1$. 

By contradiction, consider a word $w$ that does not have the form
$\paren{a_1 a_2 \ldots a_k}^{\frac{n}{k}}$ and assume that it
minimizes sum (\ref{eq1}). 
Then there exists a pair of positions $m_\ell<m_r$ such that
$w[m_\ell]=w[m_r]$ and $m_r-m_\ell<k$. Among all such pairs,
consider the one with minimal $m_r$. 

Show that for any position $m$, $k<m<m_r$, we must have
$w[m]=w[m-k]$. This is because letter $w[m]$ cannot repeat on the left
at a distance smaller than $k$, as this would contradict the
definition of $m_r$. On the other hand, the closest occurrence of
$w[m]$ to the left cannot be at a distance larger than $k$
either. Indeed, if $w[m]=w[m']$ for some $m'<m$ and $m-m'>k$ and there
is no occurrence of $w[m]$ in $w[m'+1..m-1]$, then  subword
$w[m'+1..m'+k]$ is composed of $k-1$ letters and has length $k$, and
therefore contains a letter repeated at a distance at most $k-1$. This
contradicts again the definition of $m_r$. 

By the above, we can assume that 
$w[1..m_r-1]=(a_1..a_k)^qa_1..a_i$ (up to a permutation of alphabet
letters), $q\geq 1$, and 
$w[m_r]=a_j$ for some
$j\neq i'$ where $i'=i+1$ if $i<k$ and $i'=1$ if $i=k$. 
Consider the
closest position of $a_{i'}$ to the right of $m_r$, that we denote
$m'$. (If such a position does not exist, the proof below will
trivially apply.) 

We modify $w$ by simultaneously
\begin{itemize}
\item replacing all occurrences of $a_j$ at positions $\geq m_r$ by
  $a_{i'}$, and
\item replacing all occurrences of $a_{i'}$ at positions $\geq m'$ by
  $a_{j}$.
\end{itemize}
We show that this modification makes sum (\ref{eq1}) smaller. 

The only distances between consecutive occurrences of letters that
will be affected by the modification of $w$ are the distance
$m_r-m_\ell$ between the corresponding occurrences of $a_j$ and the
distance $m'-(m_r-k)$ between the occurrences of $a_{i'}$. The new
distances become respectively $k$ (between occurrences $m_r$ and $m_r-k$ of $a_{i'}$) and
$m'-m_l$ (between corresponding occurrences of $a_j$). We show that 
\[
\frac{1}{m_r-m_\ell}+\frac{1}{m'-(m_r-k)}>\frac{1}{k}+\frac{1}{m'-m_l}.\]
This will show that sum (\ref{eq1}) becomes
smaller after the modification. For this, we show that 
\[
\frac{1}{m_r-m_\ell}-\frac{1}{k}>\frac{1}{m'-m_l}+\frac{1}{m'-(m_r-k)},\]
or
\[\frac{k-m_r+m_\ell}{(m_r-m_\ell)k}>\frac{k-m_r+m_\ell}{(m'-m_l)(m'-(m_r-k))}.\]
The numerators of both sides are equal. In denominator, we have
$m'-m_l>m_r-m_\ell$ and $m'-(m_r-k)>k$, which proves the inequality. 

We obtained a contradiction with the assumption that $w$ minimizes sum
(\ref{eq1}). This shows that a word that minimizes sum (\ref{eq1})
must have the form $w=\paren{a_1 a_2 \ldots a_k}^{\frac{n}{k}}$ (modulo a
permutation of alphabet letters). On this word, sum (\ref{main-sum})
and sum (\ref{eq1}) are both equal $\frac{n}{k}-1$. This proves that
$w$ also minimizes sum (\ref{main-sum}). 
\end{proof}

\section{Words with repetitions of bounded period}
\label{boundedp}

In this section, we study sum (\ref{main-sum}) in the case when all
repetitions in $w$ are of period at most $p$. Recall that $k$ is the
alphabet size. 

\begin{theorem}
\label{t4}
Let the period of all repetitions of a word $w$ ($|w|=n$) be bounded
by $p$. Then
$\sum_{r\in \mr{w}}\paren{e(r)-1}\leq n+3kp(\ln(p)+1)$.
\end{theorem}

The proof will use the Fine and Wilf's theorem (see
e.g. \cite{Lothaire83}) asserting that if $w$ have (not necessarily
minimal) periods $p_1$ and $p_2$ and $|w|\geq p_1+p_2-gcd(p_1,p_2)$, 
then $w$ has also the period $gcd(p_1,p_2)$. This implies, in particular, that two
different repetitions with minimal periods $p_1$ and $p_2$ cannot intersect on
$(p_1+p_2)$ letters or more. 

\begin{proof}
Consider a word $w$ such that the period of any repetition in $w$ is
bounded by $p$. 

Assume that for some letter $a$, two occurrences of $a$ are located at
a distance $3p$ or more. 
%
Consider a repetition $r$ defined by the match of these two
occurrences of $a$. 
We will show that $r$ has a very particular form, namely
\begin{itemize}
\item[(a)] all letters within a root of $r$ are different,
\item[(b)] any letter of $r$ does not occur outside $r$.
\end{itemize}

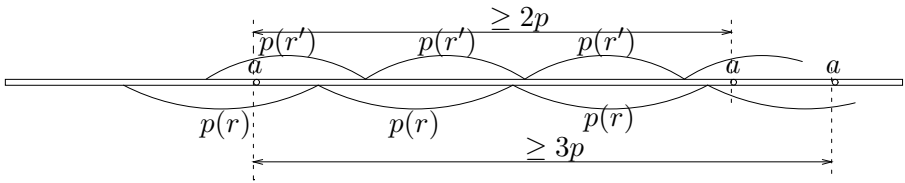
\begin{figure}[h]
\begin{center}
\setlength{\unitlength}{0.00041667in}
\begingroup\makeatletter\ifx\SetFigFont\undefined%
\gdef\SetFigFont#1#2#3#4#5{%
  \reset@font\fontsize{#1}{#2pt}%
  \fontfamily{#3}\fontseries{#4}\fontshape{#5}%
  \selectfont}%
\fi\endgroup%
{\renewcommand{\dashlinestretch}{30}
\begin{picture}(11424,2208)(0,-10)
\put(6612,312){\makebox(0,0)[lb]{\smash{{\SetFigFont{9}{10.8}{\familydefault}{\mddefault}{\updefault}$\geq 3p$}}}}
\put(2749.500,3593.250){\arc{5367.218}{1.0915}{2.0501}}
\put(7699.500,3593.250){\arc{5367.218}{1.0915}{2.0501}}
\put(10173.137,3588.136){\arc{5356.886}{1.3300}{2.0505}}
\put(5599.500,-269.250){\arc{3713.257}{4.1356}{5.2892}}
\put(7624.500,-269.250){\arc{3713.257}{4.1356}{5.2892}}
\put(3574.500,-269.250){\arc{3713.257}{4.1356}{5.2892}}
\put(9623.842,-179.447){\arc{3535.152}{4.1201}{5.0069}}
\put(3204,1251){\ellipse{72}{72}}
\put(10554,1251){\ellipse{72}{72}}
\put(9264,1251){\ellipse{72}{72}}
\dashline{60.000}(10512,1437)(10512,87)
\path(3282.000,267.000)(3162.000,237.000)(3282.000,207.000)
\path(3162,237)(10512,237)
\path(10392.000,207.000)(10512.000,237.000)(10392.000,267.000)
\dashline{60.000}(3162,2037)(3162,1662)(3162,12)
	(3162,12)(3162,12)(3162,12)
\path(3282.000,1917.000)(3162.000,1887.000)(3282.000,1857.000)
\path(3162,1887)(9237,1887)
\path(9117.000,1857.000)(9237.000,1887.000)(9117.000,1917.000)
\path(12,1287)(11412,1287)(11412,1212)
	(12,1212)(12,1287)
\dashline{60.000}(9237,2037)(9237,987)
\put(3087,1362){\makebox(0,0)[lb]{\smash{{\SetFigFont{9}{10.8}{\familydefault}{\mddefault}{\updefault}$a$}}}}
\put(6162,1962){\makebox(0,0)[lb]{\smash{{\SetFigFont{9}{10.8}{\familydefault}{\mddefault}{\updefault}$\geq 2p$}}}}
\put(3237,1662){\makebox(0,0)[lb]{\smash{{\SetFigFont{9}{10.8}{\familydefault}{\mddefault}{\updefault}$p(r')$}}}}
\put(5262,1662){\makebox(0,0)[lb]{\smash{{\SetFigFont{9}{10.8}{\familydefault}{\mddefault}{\updefault}$p(r')$}}}}
\put(7287,1662){\makebox(0,0)[lb]{\smash{{\SetFigFont{9}{10.8}{\familydefault}{\mddefault}{\updefault}$p(r')$}}}}
\put(9162,1362){\makebox(0,0)[lb]{\smash{{\SetFigFont{9}{10.8}{\familydefault}{\mddefault}{\updefault}$a$}}}}
\put(10437,1362){\makebox(0,0)[lb]{\smash{{\SetFigFont{9}{10.8}{\familydefault}{\mddefault}{\updefault}$a$}}}}
\put(2487,612){\makebox(0,0)[lb]{\smash{{\SetFigFont{9}{10.8}{\familydefault}{\mddefault}{\updefault}$p(r)$}}}}
\put(4887,612){\makebox(0,0)[lb]{\smash{{\SetFigFont{9}{10.8}{\familydefault}{\mddefault}{\updefault}$p(r)$}}}}
\put(7362,687){\makebox(0,0)[lb]{\smash{{\SetFigFont{9}{10.8}{\familydefault}{\mddefault}{\updefault}$p(r)$}}}}
\put(5224.500,3593.250){\arc{5367.218}{1.0915}{2.0501}}
\end{picture}
}
\end{center}
\caption{Proof of condition (a) of Theorem~\ref{t4}}
\label{f2}
\end{figure}
First observe that since the period of $r$ cannot exceed $p$, then
the two occurrences of $a$ are separated by at least three
periods $\p{r}$. 
To prove (a), assume that there is another occurrence of $a$ in the suffix root
of $r$ 
(cf Figure~\ref{f2}). Then, there is a repetition $r'$
formed by matching this occurrence of $a$ with the left occurrence
of $a$. These two occurrences are separated by $3p-\p{r}\geq 2p$
letters. Consider $\p{r'}$. Since $\p{r'}\leq p$,
there are at least $2\p{r'}$ letters between these two occurrences of
$a$. This means that repetitions $r$ and $r'$ intersect by length at least
$2\cdot \max\{\p{r},\p{r'}\}$ and by Fine and Wilf's theorem, $r$ and $r'$
must coincide. This contradiction proves that $a$ cannot have another
occurrence within a root of $r$. More generally, the same argument
shows that any letter occurs in a root only once. 

Condition (b) is proved by a similar argument. Assume that some letter
$b$ of $r$ occurs outside $r$, for instance to the right of $r$. Then
consider the match of this occurrence of $b$ with the leftmost
occurrence of $b$ inside $r$. This match defines a repetition
$r'$. Similar to part (a), $r$ and $r'$ intersect by length at least
$2\cdot \max\{\p{r},\p{r'}\}$ and therefore must coincide by Fine and Wilf's
theorem. This contradicts to the assumption that of an occurrence of
$b$ outside $r$ and proves (b). 

Now, we split all repetitions into two disjoint classes: repetitions
verifying conditions (a) and (b) and the others, called
respectively repetitions of type 1 and repetitions of type 2. By
condition (b), for any word $w$, repetitions of type 1 and type 2 in
$w$ are non-intersecting. Furthermore, conditions (a) and (b) insure that two distinct repetitions of type 1
cannot intersect. Therefore, all repetitions of type 1 together cannot
contribute more than $n$ to the sum. 

On the other hand, repetitions of type 2 cannot
take more than $3kp$ letters altogether in $w$, as each letter cannot occur more
than $3p$ times as this would lead to a repetition of type 1 by the
above reasoning. Therefore, by Corollary~\ref{c1}, sum
(\ref{main-sum}) for repetitions of type 2 is bounded by $3kp(\ln(p)+1)$. This
gives the final bound $n+3kp(\ln(p)+1)$. 
\end{proof}

Notice that the bound in Theorem~\ref{t4} is optimal in some sense, since
sum (\ref{main-sum}) is $n-1$ for the word $a^n$ and $\Theta\paren{kp\ln(p)}$ for the word
$(a_1a_1a_2a_2)^{p/4}(a_3a_3a_4a_4)^{p/4}\ldots$, according to
Theorem~\ref{t2}. 

\section{Concluding remarks}

Many questions related to the combinatorics of repetitions of
arbitrary exponent remain unanswered. A major such question is the
precise bound on the number of such repetitions. Corollary~\ref{c2}
provides an $O(n\log n)$ bound for the exponents at least
$(1+\varepsilon)$, for any fixed $\varepsilon>0$. It would be of great
interest to refine this bound, possibly depending on $\varepsilon$. It
is not excluded that, possibly starting from some $\varepsilon>0$, or
even for any fixed $\varepsilon>0$, the number of all repetitions of
exponent at least $(1+\varepsilon)$ is $O(n)$. This is a challenging
question, that seems, however, difficult to solve, as it would
generalize the result of
\cite{KolpakovKucherovFCT99,KolpakovKucherovJDA00} on the linear
number of runs.

\bibliography{biblio}

\end{document}